\documentclass[10pt,journal,final,twocolumn]{IEEEtran}

\usepackage{graphicx}
\usepackage{subfigure}
\usepackage{epstopdf}
\usepackage{epsfig}
\usepackage[cmex10]{amsmath}
\usepackage{amssymb}
\usepackage{times}
\usepackage{ntheorem}
\usepackage{pifont}
\usepackage{stfloats}
\usepackage{bm}
\usepackage{color}
\usepackage{makecell}
\usepackage{array}
\usepackage{algorithm}
\usepackage{algorithmic}
\usepackage{multirow}
\usepackage{amsmath}
\usepackage{xcolor}
\usepackage{url}

\begin{document}
%
\title{Reinforcement Learning Based Cooperative Coded Caching under Dynamic Popularities in Ultra-Dense Networks}

\author{\IEEEauthorblockN{Shen Gao, Peihao Dong,~\IEEEmembership{Student Member,~IEEE}, Zhiwen Pan,~\IEEEmembership{Member,~IEEE},\\
and Geoffrey Ye Li,~\IEEEmembership{Fellow,~IEEE}}

\thanks{
The work of S. Gao and Z. Pan was supported by National Key Research and Development Project under Grant 2018YFB1802402 and 333 Program of Jiangsu under Grant BRA2017366. The work of S. Gao was also supported by China Scholarship Council (CSC) under Grant 201706090065. \emph{(Corresponding author: Zhiwen Pan, Peihao Dong.)}

S. Gao and Z. Pan are with the National Mobile Communications Research Laboratory, Southeast University, Nanjing, Jiangsu 210096, China, and are also with the Purple Mountain Laboratories, Nanjing, Jiangsu 211100, China (e-mail: gaoshen@seu.edu.cn; pzw@seu.edu.cn).

P. Dong is with the National Mobile Communications Research Laboratory, Southeast University, Nanjing, Jiangsu 210096, China (e-mail: phdong@seu.edu.cn).

G. Y. Li is with the School of Electrical and Computer Engineering, Georgia Institute of Technology, Atlanta, GA 30332 USA (e-mail: liye@ece.gatech.edu).
}
}

\IEEEtitleabstractindextext{%
\begin{abstract}
For ultra-dense networks with wireless backhaul, caching strategy at small base stations (SBSs), usually with limited storage, is critical to meet massive high data rate requests. Since the content popularity profile varies with time in an unknown way, we exploit reinforcement learning (RL) to design a cooperative caching strategy with maximum-distance separable (MDS) coding. We model the MDS coding based cooperative caching as a Markov decision process to capture the popularity dynamics and maximize the long-term expected cumulative traffic load served directly by the SBSs without accessing the macro base station. For the formulated problem, we first find the optimal solution for a small-scale system by embedding the cooperative MDS coding into Q-learning. To cope with the large-scale case, we approximate the state-action value function heuristically. The approximated function includes only a small number of learnable parameters and enables us to propose a fast and efficient action-selection approach, which dramatically reduces the complexity. Numerical results verify the optimality/near-optimality of the proposed RL based algorithms and show the superiority compared with the baseline schemes. They also exhibit good robustness to different environments.
\end{abstract}

}

\maketitle

\IEEEdisplaynontitleabstractindextext

\IEEEpeerreviewmaketitle

\section{Introduction}

\IEEEPARstart{I}{n} recent years, the mobile data traffic increases dramatically and overwhelms the current fourth generation (4G) systems \cite{J. G. Andrews_1}. As one of the key technologies in the fifth generation (5G) systems, ultra-dense network (UDN) can improve the system throughput significantly by deploying multiple small base stations (SBSs) that coexist with the macro base stations (MBSs) \cite{J. Hoadley}, \cite{H. Dhillon}. By exploiting efficient interference coordination \cite{ITU-R}, UDN can improve the system throughput per unit area (on the scale of $\textrm{km}^{2}$) almost linearly with the number of SBSs. Wireless backhaul technology \cite{smallcell}, \cite{X. Ge} is regarded as a feasible solution to overcome the installation obstacle to wired backhaul caused by the expensive costs and hard-to-reach locations of SBSs. However, the need to forward massive data traffic poses as a main challenge for wireless backhaul due to its limited spectrum resources. One thus has to resort to smart content caching at the edge of the network to alleviate backhaul congestion so as to afford satisfactory quality of experience (QoE) \cite{X. Wang}, \cite{A. Liu}. In \cite{Intel}, it is shown that the data traffic can be decreased by $45\%$ through proper caching. Nonetheless, the limited storage capacity at each SBS crucially calls for proper prioritization of content for caching at the cell edge to best meet user requests. How to design efficient caching strategies has thus drawn much research interest. Recently, various trials driven by machine learning are widely emerged to optimize communication networks and have achieved great success \cite{F. Tang}$-$\cite{L. Liang_b}. Therefore, reinforcement learning (RL) may be a promising solution for caching design compared to the traditional optimization based methods.

\subsection{Related work}

The authors in \cite{K. Poularakis1}$-$\cite{C. Yang} have designed the non-cooperative caching strategy, where the user fetches the requested full content from the BS. A joint design of caching and routing policies in \cite{K. Poularakis1} is formulated under hard bandwidth constraints of the SBSs and optimized based on the facility locations. To cope with the limited cache space, the caching strategy in \cite{K. Poularakis2} is optimized based on multicast transmission by using randomized-rounding techniques. Based on diversity transmission, a probabilistic caching strategy is proposed in \cite{Y. Zhou} to minimize the content delivery latency in spatially clustered networks. In \cite{Y. Cui}, a joint caching and multicasting approach is designed for large-scale heterogeneous networks (HetNet), based on which the successful transmission probability is derived and optimized to provide the best performance. In \cite{Ejder}, the average delivery rate is analyzed for a two-tier HetNet with inter-tier and intra-tier dependence, where the most popular contents are cached at the SBSs. The tier-level content placement is investigated in \cite{J. Wen} to maximize the hit probability for a multi-tier HetNet. A cache-based content delivery scheme is proposed in \cite{C. Yang} in terms of the ergodic rate, outage probability, throughput, and delay for a three-tier HetNet including BSs, relays, and device-to-device pairs.

In contrast to the non-cooperative caching strategy simply storing full contents, it is preferable for the BSs to store the fragments of contents and serve a common user in a cooperative manner, which reduces the backhaul overhead significantly but at the cost of relatively high computational complexity. In \cite{K. Shanmugam}, a coded scheme is compared with a non-cooperative scheme and is shown to be superior in terms of the expected downloading delay. Two fundamental metrics, the expected backhaul rate and the energy consumption, are minimized in \cite{F. Gabry} for maximum-distance separable (MDS) coding based cooperative caching scheme in a two-tier HetNet. In \cite{Z. Chen}, a combined caching scheme is developed, where a reserved part of the cache space is used for cooperative caching at the SBSs.

The above work utilizes stochastic geometry and conventional optimization algorithms to optimize caching strategy based on the acquired popularity profile, which may cause performance degradation if the popularity profile evolves unpredictably with time. Fortunately, RL has been shown to perform well in sequential decision making by capturing the unknown and nondeterministic environmental dynamics for use in various caching problems in wireless networks. To minimize the long-term average energy cost, a threshold-based proactive caching strategy is proposed in \cite{S. O. Somuyiwa} for cache-enabled mobile users, where RL algorithms are used to optimize the parameters representing the threshold values. In \cite{E. Rezaei}, the cooperative transmission among local caches is exploited to achieve a trade-off between the hit ratio and the average delay based on a multi-agent RL framework, where a user will resort to other local caches when the content is unavailable at its associated local cache. The grouped linear model is introduced in \cite{N. Zhang} to obtain the predicted content requests, based on which the cache replacement is optimized by using RL with model-free acceleration. In \cite{A. Sadeghi}, a RL framework is proposed to obtain the optimal caching strategy at SBSs taking into account the space-time dynamics of the content popularity. In \cite{Y. Wei}, the probabilistic caching strategy, resource allocation, and computation offloading at fog nodes are jointly considered to minimize the average transmission delay exploiting deep RL. In \cite{S. O. Somuyiwa}$-$\cite{Y. Wei}, only full contents, instead of the fragments, are stored. As we can see in our subsequent work in this article, the power of RL can be sufficiently exploited by storing only part of contents.

\subsection{Motivation and Contribution}

In \cite{K. Poularakis1}$-$\cite{Z. Chen}, caching strategies are designed by assuming a time-invariant content popularity profile, which may not yield desirable performance when facing the dynamic popularity profile in the real scenario. In contrast, we exploit RL to capture the dynamics of content popularity evolution through interactions with the environment. The RL based solution incorporates the inherent characteristics of the content popularity profile and popularity transition and thus is suitable for the practical scenario with changeable popularity profile. On the other hand, different from full content caching in the traditional network architecture \cite{S. O. Somuyiwa}$-$\cite{Y. Wei}, we introduce the cooperative caching strategy storing coded fragments at SBSs so that the limited storage can be utilized more efficiently for UDN. The feasibility is guaranteed by the fact that the UDN architecture enables a user to be served by more than one SBS. Moreover, the significantly increasing SBSs in UDN can contribute much more unbiased data to the global popularity to help the MBS improve the RL policy and to enhance the effectiveness of the cooperative coded caching. Although the cooperative coded caching is more efficient, it poses daunting challenges when formulating the RL based problem in such complex UDN. In this paper, we design the cooperative coded caching strategy for UDN exploiting RL and the main contributions can be summarized as follows.

\begin{itemize}[\IEEEsetlabelwidth{Z}]
\item[1)] Prior works either assume the time-invariant content popularity profile or use the less efficient full content caching. We develop a cooperative coded caching strategy using RL algorithms, which is a promising solution for backhaul offloading in UDN. Under an unknown dynamic environment on usage traffic, we figure out how to cooperate among the SBSs and what kind of coding method to choose to stimulate the potential of the SBS cooperation most. Afterwards, we abstract the MDS coding based cooperative caching strategy design into a RL based sequential decision making problem, which is mathematically modeled as a Markov decision process (MDP) to maximize the long-term expected cumulative traffic load served directly by the SBSs without accessing the MBS. Our formulated problem matches the real scenario well and can reduce the performance loss caused by popularity profile mismatch.

\item[2)] For the formulated RL based problem, we successfully find the matched solution for performance maximization. After evaluating the feasibility, performance, and complexity of the possible RL algorithms comprehensively, we propose to embed the complicated cooperative MDS coding into the Q-learning based solution. The developed Q-learning based cooperative coded caching strategy is shown optimal for the formulated RL problem with an acceptable complexity in the small-scale system.

\item[3)] As the system dimensionality becomes large, the Q-learning based algorithm will malfunction and finding the appropriate solution is challenging due to the prohibitively huge action space. To address this problem, we first approximate the state-action value function heuristically according to the instantaneous reward. The approximated state-action value function includes only a small number of learnable parameters and enables us to propose a fast and efficient action-selection approach, which dramatically reduces the complexity. The developed value function approximation based algorithm can be flexibly applied to the large-scale system yet still yielding near-optimal performance.
\end{itemize}

The rest of the paper is organized as follows. The considered system model is described in Section II. Section III presents the RL based problem formulation and proposes a Q-learning based solution for a small-scale cooperative coded caching system. The large-scale solution is further developed in Section IV by using value function approximation. Numerical results are presented in Section V to verify our proposed solutions. Finally, Section VI provides concluding remarks.

\section{System Model}

In this section, we first give an overview of the network model, followed by the description of the content popularity profile. Then we introduce the MDS coded cooperative caching, based on which the cooperative transmission among BSs is provided.

\subsection{Network Model}

\begin{figure}[!t]
\centering
\includegraphics[trim=15 15 0 10, width=2.8in]{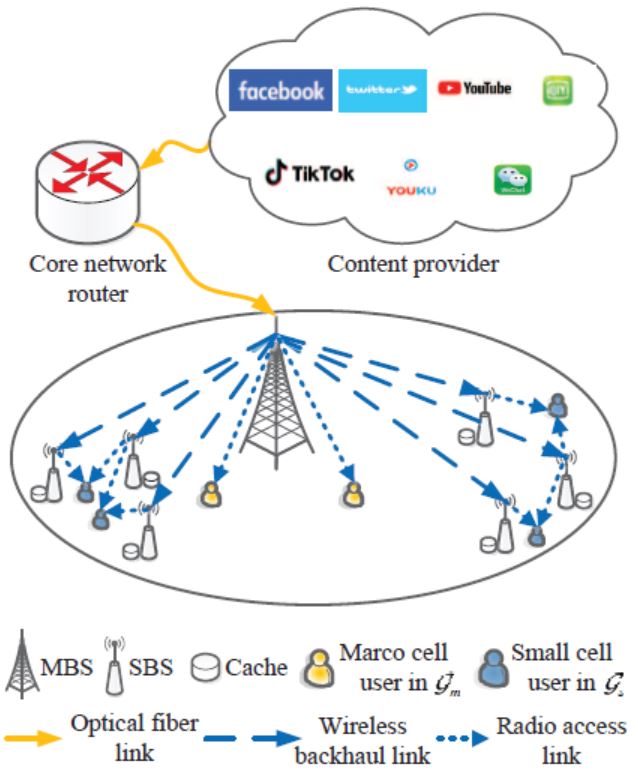}
\caption{Network model.}\label{system_model}
\end{figure}

Consider the downlink of a two-tier UDN with an MBS and $p$ SBSs as illustrated in Fig.~\ref{system_model}. In the coverage area of the MBS, $p$ SBSs are deployed to provide better coverage and enhance system throughput. $\mathcal{G}=\{\mathcal{G}_{m},\mathcal{G}_{s}\}$ is the set of users in the network, where $\mathcal{G}_{m}$ and $\mathcal{G}_{s}$ represent the sub-sets including users that are served by the MBS directly and that connect to the MBS through a SBS, respectively, and can be specified by load balancing \cite{Q. Ye}. The links between users and BSs, i.e., MBS and SBSs, are called radio access links and those connecting SBSs and the MBS are referred to as wireless backhaul links. In addition, the MBS can retrieve all contents from the content provider through the core network router and optical fibers. The radio access links and wireless backhaul links operate on orthogonal spectra, that is, the whole spectrum $B_w$ is divided into two parts denoted as $B_{al}$ and $B_w-B_{al}$ for the radio access links and the wireless backhaul links, respectively, as in \cite{H. S. Dhillon}, \cite{H. H. Yang}. Densely-deployed SBSs bring a daunting challenge to wireless backhaul links when serving massive concurrent user requests. Users may suffer from unbearable delays, which result in a bad user experience due to the congestion of wireless backhaul links during the peak hours of transmission. Therefore, caches, which are usually with limited storage capacity, are deployed at SBSs to offload the wireless backhaul load and reduce the queueing delay so that QoE can be improved for users. The SBSs cache contents to serve the users in $\mathcal{G}_{s}$, which consequently will be focused on hereinafter. Without loss of generality, we assume that no transmission error occurs when any SBS delivers contents to the users in its coverage range. Since the users in $\mathcal{G}_{s}$ can only be served by the SBSs, the transmission error will impact the performance similarly no matter whether the SBSs cache the contents or not.

The considered heterogeneous UDN model abstracted from \cite{Nokia} has been recognized as the solution for the future network architecture and widely investigated in lots of studies \cite{X. Ge2}$-$\cite{W. Wen}. It is noted that except the well-known effect on improving the system throughput significantly, the densely-deployed SBSs can also facilitate the RL based cooperative coded caching design. Specifically, the densely-deployed SBSs can collect and report sufficient user requests to the MBS. Only in this case, the user requests received by the MBS can reflect the actual user request distribution unbiasedly and accurately according to the law of large numbers. Then the MBS is able to design a more efficient caching strategy that works well in the real scenarios by using RL algorithms. Furthermore, the high density of SBSs means that a user can be served by more SBSs and thus UDN is more favorable for the cooperative coded caching compared to the traditional network architecture.

\subsection{Content Popularity Profile}

Users in the network request contents from a set denoted by $\mathcal{C}=\{1,2,\ldots,C\}$, where the size of each content is equal to $B$ \cite{F. Gabry}.\footnote{Note that, the approach in this paper can also be applied to the case of different content sizes. Specifically, contents with different sizes can be properly combined or divided into pieces with the same size to guarantee the equal size for each content.} According to the alternation of peak and off-peak hours, the whole transmission time is slotted and each time slot $t$ contains peak hours and off-peak hours of transmission.\footnote{In a time slot, peak hours refer to the period when the network experiences high traffic load with numerous requests generated. The following off-peak hours represent the period when the content requests from users are relatively inactive. The duration of peak hours is usually longer than that of off-peak hours. In the practical scenario, the network operator may adjust the length of the time slots and the durations of peak and off-peak hours dynamically according to the traffic load.} Denote $\mathbf{N}_{i}(t)=[N_{i1}(t),N_{i2}(t),\ldots,N_{iC}(t)]$ as the vector composed of the number of user requests for each content during the peak hours of time slot $t$ collected by the $i$th SBS for $i=1,2,\ldots,p$. $\mathbf{N}(t)=[N_{1}(t),N_{2}(t),\ldots,N_{C}(t)]$ is the aggregation of $\mathbf{N}_{1}(t),\mathbf{N}_{2}(t),\ldots,\mathbf{N}_{p}(t)$ at the MBS. $\boldsymbol{\theta}(t)=[\theta_{1}(t),\theta_{2}(t),\ldots,\theta_{C}(t)]$ denotes the corresponding content popularity vector, defined as
\begin{equation}
\label{eqn_theta}
\theta_{c}(t)=\frac{N_{c}(t)}{\sum_{j=1}^C N_{j}(t)},\quad c\in\mathcal{C}.
\end{equation}

The user requests used to compute the content popularity profile exhibit the specific distribution. The user request distribution has been studied in \cite{L. Breslau} and \cite{Engin} by using the real traces of user requests and modeled as a Zipf-like distribution. Specifically, six traces of the web proxies are collected in \cite{L. Breslau} from academic, corporate, and Internet service provider environments and prove to follow Zipf-like behaviors. Authors in \cite{Engin} investigate the popular wireless mobile network based on a mass of representative data from the telecom operator and demonstrate the good fitness of the Zipf-like distribution. It is noted that users in an area generally exhibit diverse individual preferences for the contents while the overall user requests of this area aggregating the individual preference of each user follow the Zipf-like distribution. According to \cite{S. Gitzenis}, the skewness of the Zipf-like distribution is dependent on the specific application. In brief, Zipf-like distribution is able to well depict the real user request distribution of various networks and thus is widely adopted in \cite{K. Poularakis1}, \cite{Y. Cui}, \cite{K. Shanmugam}, \cite{F. Gabry}, \cite{A. Sadeghi}. In this paper, we consider the mobile UDN and thus Zipf-like distribution can be safely used to model the user request distribution. As it will be mentioned in Section V, the simulation data are generated following Zipf-like distribution with the skewness set according to \cite{S. Gitzenis}.\footnote{It is noted that the RL based approach is designed in a model-free manner and its ability to capture the inherent popularity evolution is independent of any specific distribution model according to Section III and IV. To verify the effectiveness of the proposed approach, we employ the practical Zipf-like distribution to generate the simulation data.}

\subsection{MDS Coded Cooperative Caching}

From \cite{K. Shanmugam}, storing fragments of contents, instead of complete contents, in caches actually performs better in offloading the wireless backhaul. In the case of storing uncoded fragments, a request from a user will be met only when the user collects all content fragments. In contrast to simply splitting a content into multiple coarse fragments, utilizing coded caching scheme can further improve the performance. For MDS coding, $B$ information bits are coded as a string of parity bits with an arbitrary length and then split into multiple packets. According to \cite{K. Shanmugam}, a user can recover the full content so long as it collects a certain number of packets with the total size no less than $B$ parity bits since the individual identity of each bit is irrelevant. Thus, caching design at the SBSs matters since it determines how many parity bits the SBSs can deliver directly to the user without resorting to the MBS. A cache memory with a limited size of $K\times B$ bits is installed at each SBS. Let $\mathbf{a}(t)=[a_{1}(t),a_{2}(t),\ldots,a_{C}(t)]$ denote the SBSs caching decision vector designed by the MBS during the off-peak hours of time slot $t$, where $a_{c}(t)\in[0,1]$ is the normalized fraction of content $c$ cached at the SBSs. The elements in $\mathbf{a}(t)$ are subject to the constraint $\sum_{i=1}^C a_{i}(t)=K$ and the set of stored contents with $a_{c}(t)\geq a_{c}(t-1)$ is denoted by $\mathcal{C}^{'}(t)$. The MBS converts the content $c\,(c\in\mathcal{C}^{'}(t))$ into $B_{c}^{\textrm{MDS}}$ parity bits expressed as
\begin{equation}
\label{eqn_BMDS}
B_{c}^{\textrm{MDS}}=(p+1)B,\quad c\in\mathcal{C}^{'}(t).
\end{equation}
The $B_{c}^{\textrm{MDS}}$ parity bits are divided into two non-overlapped candidate sets, $\mathcal{B}_{c}^{\textrm{SBS}}$ and $\mathcal{B}_{c}^{\textrm{MBS}}$, respectively. $\mathcal{B}_{c}^{\textrm{SBS}}$ contains $pB$ bits and is equally divided into $p$ non-overlapped candidate sets, $\mathcal{B}_{c}^{1},\ldots,\mathcal{B}_{c}^{p}$, for $p$ SBSs, respectively. $\mathcal{B}_{c}^{\textrm{MBS}}$ including the remaining $B$ bits is the candidate set for the MBS.\footnote{$\mathcal{B}_{c}^{\textrm{MBS}}$, $\mathcal{B}_{c}^{\textrm{SBS}}$, and $\mathcal{B}_{c}^{1},\ldots,\mathcal{B}_{c}^{p}$ are constant for all time slots.} Then the MBS delivers $(a_{c}(t)-a_{c}(t-1))B$ bits from each of $\mathcal{B}_{c}^{1},\ldots,\mathcal{B}_{c}^{p}$ to the corresponding SBS for caching. Assuming there are $d$ SBSs serving the user simultaneously, the MBS should preserve arbitrary $(1-da_{c}(t))B$ bits from $\mathcal{B}_{c}^{\textrm{MBS}}$ when $da_{c}(t)\leq1$.

\subsection{BSs Cooperative Transmission Scheme}

\begin{figure}[!t]
\centering
\includegraphics[trim=15 15 0 10, width=3.2in]{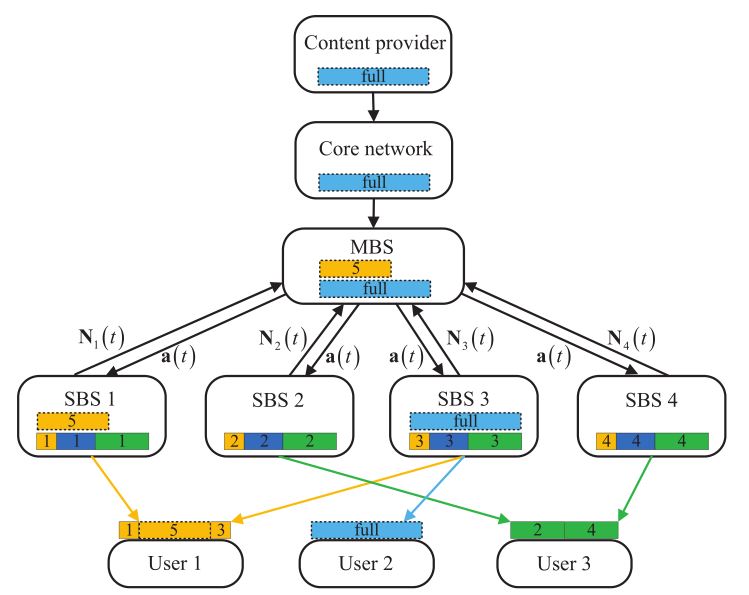}
\caption{Hierarchical framework of cooperative coded caching at the cell edge. Different contents are distinguished by different colors. The indexed rectangles with the same color denote the different coded fragments of a content, where the solid and dashed lines indicate that the fragments are stored in the SBSs and the MBS, respectively. The rectangle with the dashed line and label ``full" represents the full content unavailable at any of SBSs that can cover the user and fetched from the content provider.}\label{cooperative_transmission}
\end{figure}

In UDN, densely-deployed SBSs enable users to be served by multiple SBSs simultaneously. Moreover, MDS coded caching at each SBS makes the cooperative transmission more effective. User $g\in\mathcal{G}_{s}$ selects $d$ SBSs providing the strongest reference signal received power (RSRP) to form its cooperative serving set $\mathcal{S}_{g}^{d}$. When user $g$ requests content $c$ during the peak hours of time slot $t+1$ and $a_{c}(t)\neq0$, it will receive $da_{c}(t)B$ coded bits from $d$ associated SBSs. If $da_{c}(t)\geq1$, the user is able to restore the full content $c$ directly. Otherwise, the MBS chooses a SBS $s^{*}$ with the strongest RSRP from $\mathcal{S}_{g}^{d}$ and sends the complementary $(1-da_{c}(t))B$ bits to $s^{*}$. Then $s^{*}$ transmits these coded bits to user $g$. In the case of $a_{c}(t)=0$, the MBS transmits the original content with the size of $B$ bits to $g$ via the SBS $s^{*}$.

To provide a more comprehensive illustration on how this network works, Fig.~\ref{cooperative_transmission} shows the hierarchical framework of cooperative coded caching abstracted from Fig.~\ref{system_model}. All SBSs report their respective observations on the numbers of content requests to the MBS and the latter makes caching decision for them, which leads to three possible service modes for different content requests. User 3 receives coded fragments 2 and 4, both with the size of $\frac{B}{2}$ bits of the desired content from two different SBSs, and then performs recovery. In contrast, user 1 only collects coded fragments 1 and 3 with the size of $\frac{B}{6}$ bits, which are insufficient to recover the desired content. To fill the rest of the content, the MBS sends the complementary coded fragment 5 with the size of $\frac{2B}{3}$ bits to user 1 through SBS 1. In addition, user 2 requests a content unavailable at any of SBSs that can cover it, i.e., the rectangle with the label ``full" in Fig.~\ref{cooperative_transmission}, and the MBS sends the full content with the size of $B$ bits to user 2 through SBS 3.

\section{Reinforcement Learning based Cooperative Coded Caching}

In this section, the elements in the transition tuple of MDP are first specified in the cooperative coded caching scenario, based on which the corresponding objective function is formulated. Then a Q-learning based cooperative coded caching strategy that provides optimal solution is proposed to solve the formulated problem in a small-scale system.

\subsection{Reinforcement Learning based Problem Formulation}

\begin{figure}[!t]
\centering
\includegraphics[trim=15 15 0 10, width=3.5in]{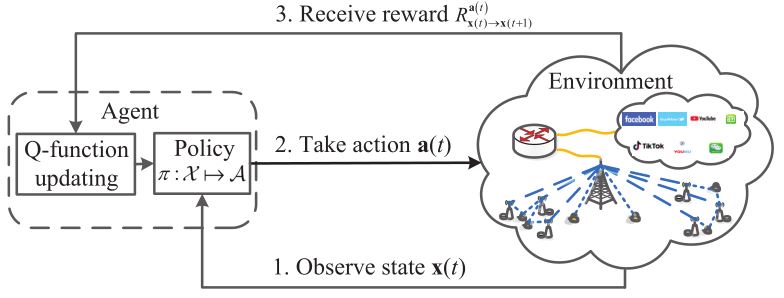}
\caption{The agent-environment interactions in a reinforcement learning problem.}\label{RL_model}
\end{figure}

As shown in Fig. 3, the agent including the MBS and the SBSs in its coverage interacts with the environment that is regarded as everything outside the agent, where the MBS makes the decision and then the SBSs execute this decision. The process can be abstracted into a RL problem, which can be modeled as a MDP and expressed as a transition tuple $\left(\mathcal{X},\mathcal{A},P_{\mathbf{x}(t)\rightarrow\mathbf{x}(t+1)}^{\mathbf{a}(t)},R_{\mathbf{x}(t)\rightarrow\mathbf{x}(t+1)}^{\mathbf{a}(t)}\right)$. At time slot $t$, the agent in the environment observes a state $\mathbf{x}(t)$, which belongs to the state space $\mathcal{X}$, and then takes an action $\mathbf{a}(t)$ chosen from the action space $\mathcal{A}$ according to the policy $\pi:\mathcal{X}\mapsto\mathcal{A}$. After that, the environment evolves from the current state $\mathbf{x}(t)$ to a new one, $\mathbf{x}(t+1)$, in a transition probability $P_{\mathbf{x}(t)\rightarrow\mathbf{x}(t+1)}^{\mathbf{a}(t)}$ and feeds back the reward $R_{\mathbf{x}(t)\rightarrow\mathbf{x}(t+1)}^{\mathbf{a}(t)}$ to the agent. The policy $\pi$ needs to be updated at each time slot until obtaining the optimal state-action value function, called Q-function. The reward received from the environment is used to update the Q-function and prepare for making decision in the following slot. To be specific, key elements of the RL based cooperative coded caching design are described as follows.

\emph{1) State: }As the characterization of the environment, the state is unknown to the agent and needs to be acquired through the observation. During the off-peak hours of time slot $t$, the observation result includes two parts: the numbers of requests for all contents during the peak hours of time slot $t$, $\mathbf{N}(t)$, and the caching decision made during the off-peak hours of time slot $t-1$, $\mathbf{a}(t-1)$, which are necessary for the selection of $\mathbf{a}(t)$. Furthermore, $\mathbf{N}(t)$ is normalized as $\boldsymbol{\theta}(t)$ according to (\ref{eqn_theta}) to acquire the underlying transition mode of content requests. Therefore, the state at time slot $t$ can be expressed as $\mathbf{x}(t)=\left[\boldsymbol{\theta}(t), \mathbf{a}(t-1)\right]$.

\emph{2) Action: }Based on the observed state $\mathbf{x}(t)$ and the updated policy $\pi$, the agent will decide which and how many contents should be stored at the SBSs during the off-peak hours of time slot $t$. Hence, the action is denoted as the caching decision vector $\mathbf{a}(t)=[a_{1}(t),a_{2}(t),\ldots,a_{C}(t)]$, where $\sum_{i=1}^Ca_{i}(t)=K$.

\emph{3) Transition Probability: }Following the action $\mathbf{a}(t)$, the environment state transits to the next state $\mathbf{x}(t+1)$ at the end of peak hours of time slot $t+1$ with probability $P_{\mathbf{x}(t)\rightarrow\mathbf{x}(t+1)}^{\mathbf{a}(t)}$. It is usually determined by the environment, independent of the agent.

\emph{4) Reward Design: }The agent will receive a reward when the environment state transits to $\mathbf{x}(t+1)$. The system performance can be enhanced when the designed reward at each time slot correlates with the desired goal. In the studied cooperative coded caching problem, our goal is to maximize the total traffic load served directly by the SBSs without accessing the MBS over all the time slots, which is consistent with the general aim of RL that maximizes the expected cumulative discounted rewards. Therefore, the received reward at time slot $t+1$ is set as
\setlength{\arraycolsep}{0.0em}
\begin{eqnarray}
\label{eqn_reward}
R_{\mathbf{x}(t)\rightarrow\mathbf{x}(t+1)}^{\mathbf{a}(t)}&&\!=\!\sum_{i=1}^C N_{i}(t\!+\!1)\!-\!p\sum_{i=1}^C\max(a_{i}(t)\!-\!a_{i}(t\!-\!1),0)\nonumber\\
&&-\!\sum_{i=1}^C N_{i}(t\!+\!1)\sum_{j=1}^C \theta_{j}(t\!+\!1)\max(1\!-\!da_{j}(t),0),\nonumber\\
&&
\end{eqnarray}
where the first term denotes the total traffic load, i.e., the total number of content requests from users, during the peak hours at time slot $t+1$, the second term represents the traffic load of updating the contents cached in the SBSs during the off-peak hours at time slot $t$, and the third term accounts for the traffic load of transmitting the complementary coded fragments from the MBS to the SBSs during the peak hours at time slot $t+1$, respectively.\footnote{The three terms in (\ref{eqn_reward}) are normalized traffic loads obtained by dividing the corresponding actual traffic load by $B$.} Please note that there are also some other signaling overheads in the agent-environment interaction process, including the overhead for reporting the user request vectors, $\mathbf{N}_{i}(t), \forall i\in\{1,\ldots,p\}$, from all SBSs to the MBS and the overhead for broadcasting the action vector, $\mathbf{a}(t)$, in reverse, where each element in $\mathbf{N}_{i}(t)$ and $\mathbf{a}(t)$ can be represented by only tens of bits or less. Thus, these signaling overheads are negligible compared to the actual traffic loads corresponding to the second and third terms in (\ref{eqn_reward}) and can be omitted safely.

\begin{figure}[!t]
\centering
\includegraphics[trim=0 0 0 0, width=2.5in]{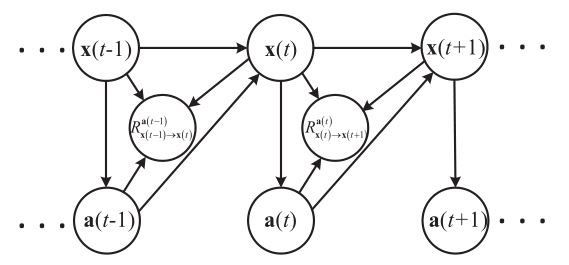}
\caption{Graphical model of the caching process.}\label{MDP}
\end{figure}

Note that it is reasonable to formulate this cooperative coded caching problem as an MDP and the feasibility is revealed by the abstracted graphical model of caching process shown in Fig.~\ref{MDP}. From the figure: i) $\mathbf{x}(t+1)$ only depends on $\mathbf{x}(t)$ but is independent of $\mathbf{x}(t-1),\mathbf{x}(t-2),\ldots$. ii) $\mathbf{a}(t)$, which is taken based on $\mathbf{x}(t)$, determines $\mathbf{x}(t+1)$ with the transition probability $P_{\mathbf{x}(t)\rightarrow\mathbf{x}(t+1)}^{\mathbf{a}(t)}$. Denoting $\mathbf{x}(t)$, $\mathbf{a}(t)$, and $\mathbf{x}(t+1)$ by $\mathbf{x}$, $\mathbf{a}$, and $\mathbf{x}^{'}$, respectively,  the transition probability satisfies $\sum_{\mathbf{x}^{'}\in\mathcal{X}} P_{\mathbf{x}\rightarrow\mathbf{x}^{'}}^{\mathbf{a}}=1$. If the MBS makes another caching decision, $\bar{\mathbf{a}}$, based on $\mathbf{x}$, then the transition probability satisfies $ P_{\mathbf{x}\rightarrow\mathbf{x}^{'}}^{\mathbf{a}}\neq P_{\mathbf{x}\rightarrow\mathbf{x}^{'}}^{\bar{\mathbf{a}}}$. iii) $R_{\mathbf{x}(t)\rightarrow\mathbf{x}(t+1)}^{\mathbf{a}(t)}$ is determined by three types of traffic loads in (\ref{eqn_reward}), which are dependent on $\mathbf{x}(t)$, $\mathbf{x}(t+1)$, and $\mathbf{a}(t)$. Therefore, the caching process satisfies the MDP properties.

\emph{5) Objective Function: }The objective of RL is to maximize the expected cumulative discounted rewards over the transition probability, which is expressed as \cite{R. S. Sutton}
\begin{equation}
\label{eqn_cumu_rewards}
G_{\gamma}^{\pi}(t)=\mathbb{E}\left[\sum_{n=0}^{\infty}\gamma^{n}R_{t+n}\right],
\end{equation}
where $\mathbb{E}\left[\cdot\right]$ represents the expectation, $\gamma\in[0,1]$ is the discounted factor, and $R_{t+n}$ denotes the successive received reward from the state $\mathbf{x}(t)$.

Use the Q-function $Q_{\gamma}^{\pi}\left(\mathbf{x},\mathbf{a}\right)$ to denote the expected cumulative discounted rewards taking an arbitrary action $\mathbf{a}$ under the current state $\mathbf{x}$ and then following the policy $\pi$. Thus, $Q_{\gamma}^{\pi}\left(\mathbf{x},\mathbf{a}\right)$ is given by
\begin{equation}
\begin{aligned}
\label{eqn_Qfunc}
Q_{\gamma}^{\pi}\left(\mathbf{x},\mathbf{a}\right)=\mathbb{E}\left[\sum_{n=0}^{\infty}\gamma^{n}R_{t+n}|\mathbf{x}(t)=\mathbf{x},\mathbf{a}(t)=\mathbf{a}\right].
\end{aligned}
\end{equation}

The Q-function of the optimal policy, $Q_{\gamma}^{\pi^{*}}\left(\mathbf{x},\mathbf{a}\right)$, can be further written in a recursive form, known as the optimal Bellman equation
\begin{equation}
\label{eqn_Qfunc2}
Q_{\gamma}^{\pi^{*}}\left(\mathbf{x},\mathbf{a}\right)=\sum_{\mathbf{x}^{'}\in\mathcal{X}}P_{\mathbf{x}\rightarrow\mathbf{x}^{'}}^{\mathbf{a}}\left(R_{\mathbf{x}\rightarrow\mathbf{x}^{'}}^{\mathbf{a}}+\gamma \max\limits_{\mathbf{a^{'}}\in\mathcal{A}}Q_{\gamma}^{\pi^{*}}\left(\mathbf{x^{'}},\mathbf{a^{'}}\right)\right),
\end{equation}
where $\mathbf{x}^{'}$ denotes the next state $\mathbf{x}(t+1)$. Once $Q_{\gamma}^{\pi^{*}}\left(\mathbf{x},\mathbf{a}\right)$ is obtained, the optimal policy can be easily determined as
\setlength{\arraycolsep}{0.0em}
\begin{eqnarray}
\label{eqn_Opt_policy}
\pi^{*}(\mathbf{x})=\arg\max\limits_{\mathbf{a}\in\mathcal{A}}Q_{\gamma}^{\pi^{*}}\left(\mathbf{x},\mathbf{a}\right).
\end{eqnarray}

In the environment with known transition tuple, the agent will learn the optimal policy not requiring real form of interactions with the environment. This can be achieved through solving the Bellman equation utilizing value iteration or policy iteration algorithm. However, it is difficult for the agent to know the tuple elements in reality, especially the transition probability and the reward function. To explore the unknown environment, Q-learning \cite{C. J. Watkins} is available to solve the Bellman equation and learn the statistics determined by the environment.

\subsection{Q-learning Algorithm for Small-Scale Solution}

In an unknown environment, the agent needs to continuously interact with the environment to improve the policy. Specifically, assuming the starting time slot for training is $t$, the agent should carry out the chosen action $\mathbf{a}(t)$ based on the current policy $\pi_{t}$ and then observe the new state $\mathbf{x}(t+1)$ and the obtained reward $R_{\mathbf{x}(t)\rightarrow\mathbf{x}(t+1)}^{\mathbf{a}(t)}$. Thereafter, the agent updates the Q-function from $Q_{t}$ to $Q_{t+1}$, and accordingly improves the policy from $\pi_{t}$ to $\pi_{t+1}$ based on
\begin{equation}
\label{eqn_Mod_policy}
\pi_{t+1}(\mathbf{x}(t))=\arg\max\limits_{\mathbf{a}\in\mathcal{A}}Q_{t+1}\left(\mathbf{x}(t),\mathbf{a}\right).
\end{equation}
When the action space is continuous, it is difficult to find the global maximum value for the non-convex function in (\ref{eqn_Mod_policy}). Hence each element in $\mathbf{a}$ is discretized into $L$ levels uniformly over the interval $[0,1]$. To determine the valid actions in $\mathcal{A}$, we first select the action vectors satisfying $\sum_{i=1}^{C} a_i(t) = K$. To reduce the complexity, we further shrink the valid action space filtrated by the first constraint by using the prior knowledge of the cooperative coded caching. That is, only the action vectors satisfying $a_i(t)\leq \frac{l_0}{L},\forall i=1,\ldots,C,$ will be selected as the final action candidates, where $l_0=\lceil\frac{L}{d}\rceil$ with $\lceil\cdot\rceil$ denoting the ceiling function. This is because taking the actions with $a_i(t)>\frac{l_0}{L},\exists i=1,\ldots,C,$ will waste the limited cache storage and thus corrupt the performance. Then the cardinality of the action space, denoted by $|\mathcal{A}|$, becomes limited. In addition, a finite number of possible popularity candidates, denoted by $|\Theta|$, is considered. Accordingly, the cardinality of the state space will be $|\mathcal{X}|=|\Theta||\mathcal{A}|$.

\begin{algorithm}
\label{alg:Q-learning}
\caption{Q-learning based Cooperative Coded Caching Strategy}
{\bf Input:}
environment simulator, action space $\mathcal{A}$\\[-0.0mm]
{\bf Output:}
the optimal caching policy $\pi^{*}(\mathbf{x})$\\[-0.0mm]
{\bf Procedure:}
\begin{algorithmic}[1]
\STATE Set the starting time slot $t=t_{0}$, the starting state $\mathbf{x}(t)=\mathbf{x}_{0}$, the discount factor $\gamma$, the step size $\lambda$, the exploration probability $\epsilon$, the initial value of Q-table: $Q_{t}(\mathbf{x},\mathbf{a})=0,\forall\mathbf{x},\mathbf{a}$, and for each state, select an arbitrary $\mathbf{a}$ as its initial action
\LOOP
\STATE (\emph{Selection of Cooperative Caching Action}): Choose a caching action based on $\varepsilon$-greedy approach during off-peak hours
\begin{displaymath}
\mathbf{a}(t)=\pi_{t}^{\epsilon}(\mathbf{x}(t))=
\begin{cases}
\pi_{t}(\mathbf{x}(t)), & \textrm{w.p.}\,1-\varepsilon\\
\forall\mathbf{a}\in\mathcal{A}, & \textrm{w.p.}\,\varepsilon
\end{cases}
\end{displaymath}
\STATE (\emph{MDS Coding}): MBS codes the contents with $a_{c}(t)\neq0$ and $a_{c}(t)\geq a_{c}(t-1)$ based on MDS coding during off-peak hours
\STATE (\emph{Coded Packets Delivery}): MBS sends $p$ different packets, each containing $(a_{c}(t)-a_{c}(t-1))B$ bits of the $c$th coded content, to $p$ SBSs correspondingly for caching during off-peak hours
\STATE (\emph{Cooperative Transmission}): User $g\in\mathcal{G}_{s}$ selects $d$ SBSs to form $\mathcal{S}_{g}^{d}$ and requests contents during peak hours of time slot $t+1$
\STATE (\emph{New State Observation}): MBS observes $\mathbf{N}(t+1)$, computes $\boldsymbol{\theta}(t+1)$ according to (\ref{eqn_theta}) during off-peak hours, and sets the new state $\mathbf{x}(t+1)=\left[\boldsymbol{\theta}(t+1),\mathbf{a}(t)\right]$
\STATE (\emph{Reward Feedback}): Environment feeds back a reward computed as (\ref{eqn_reward})
\STATE (\emph{Estimation of Cooperative Caching Action}): Estimate a caching action under the state $\mathbf{x}(t+1)$ according to
\begin{displaymath}
\mathbf{\tilde{a}}(t+1)=\arg\max\limits_{\mathbf{a}^{'}\in\mathcal{A}}Q_{t}\left(\mathbf{x}(t+1),\mathbf{a}^{'}\right)
\end{displaymath}
\STATE (\emph{Q-table Updating}): Update the Q-table:

for $\mathbf{x}=\mathbf{x}(t)$ and $\mathbf{a}=\mathbf{a}(t)$,
\begin{displaymath}
\begin{aligned}
Q_{t+1}(\mathbf{x},\mathbf{a})=&Q_{t}(\mathbf{x},\mathbf{a})+\lambda\biggl(R_{\mathbf{x}(t)\rightarrow\mathbf{x}(t+1)}^{\mathbf{a}(t)}\\
&+\gamma
Q_{t}\left(\mathbf{x}(t+1),\mathbf{\tilde{a}}(t+1)\right)-Q_{t}(\mathbf{x},\mathbf{a})\biggr)
\end{aligned}
\end{displaymath}
otherwise,
\begin{displaymath}
Q_{t+1}(\mathbf{x},\mathbf{a})=Q_{t}(\mathbf{x},\mathbf{a})
\end{displaymath}
\STATE (\emph{Policy Updating}): Update the policy:

for $\mathbf{x}=\mathbf{x}(t)$,
\begin{displaymath}
\pi_{t+1}\left(\mathbf{x}\right)=\arg\max\limits_{\mathbf{a}\in\mathcal{A}}Q_{t+1}\left(\mathbf{x},\mathbf{a}\right)
\end{displaymath}
otherwise,
\begin{displaymath}
\pi_{t+1}(\mathbf{x})=\pi_{t}(\mathbf{x})
\end{displaymath}
\STATE $t=t+1$
\ENDLOOP
\RETURN The optimal caching policy $\pi^{*}(\mathbf{x}),\forall\mathbf{x}\in\mathcal{X}$
\end{algorithmic}
\end{algorithm}

The state-action value function can be expressed by a table, i.e., Q-table, of which each element can be approximated by the average cumulative discounted rewards using the temporal difference learning and is give by
\setlength{\arraycolsep}{0.0em}
\begin{eqnarray}
\label{Q-leaning}
Q_{t+1}\left(\mathbf{x}(t),\mathbf{a}(t)\right)&&\leftarrow Q_{t}\left(\mathbf{x}(t),\mathbf{a}(t)\right)+\lambda\biggl[R_{\mathbf{x}(t)\rightarrow \mathbf{x}(t+1)}^{\mathbf{a}(t)}\nonumber\\
&&+\gamma\max\limits_{\mathbf{a}^{'}\in\mathcal{A}}Q_{t}\!\left(\mathbf{x}(t+1),\mathbf{a}^{'}\right)\!-\!Q_{t}\left(\mathbf{x}(t),\mathbf{a}(t)\right)\biggr]\!,\nonumber\\
&&
\end{eqnarray}
where $\lambda$ is step size parameter.

According to the approximation method of the state-action value in (\ref{Q-leaning}) and the updating rule of the policy in (\ref{eqn_Mod_policy}), the Q-learning algorithm based cooperative coded caching strategy is described in Algorithm 1. The data used for learning the optimal policy are generated by the agent and an environment simulator. The content requests of users are generated based on a Zipf-like distribution \cite{S. Gitzenis} by the simulator. With the selected caching decision vector of SBSs, the simulator generates the next state and the reward. The initial values in the Q-table are set as $0$. The policy used for deciding the caching decision vector is first set randomly and then is improved with the updated Q-table. In the MDP case, it has been proved in \cite{R. S. Sutton} that all the state-action values in the Q-table will converge to the optimal values with probability 1 under the assumption that all the state-action values are updated for an infinite number of times and the stochastic approximation conditions on $\lambda$. Thus, the choice of $\mathbf{a}(t)$ under state $\mathbf{x}(t)$ in (\ref{Q-leaning}) follows $\epsilon$-greedy approach to balance the exploitation and exploration and ensure the convergence. Then the optimal policy is accordingly obtained with the optimal state-action values.

In practice, the Q-learning algorithm based solution applies only to small-scale system, i.e., involving small numbers of SBSs and contents and a small cache size. That is because both $|\mathcal{A}|$ and $|\mathcal{X}|$ are related to the number of SBSs, $p$, the number of contents, $C$, the cache size, $K$, and the action discretization level, $L$, in our MDP case. In the large-scale system, large $|\mathcal{A}|$ and $|\mathcal{X}|$ make it impossible to save the Q-table with huge size. Meanwhile, it may be time-consuming to converge since many state-action pairs are seldom visited. Furthermore, the Q-learning algorithm performs ergodic search of the action space $\mathcal{A}$ in steps 9 and 11 of each iteration, which leads to high time complexity. Therefore, we develop an efficient value function approximation based algorithm for the large-scale system in the next section.

\section{Value Function Approximation for Large-Scale Solution}

In this section, we develop a value function approximation algorithm based solution for large-scale cooperative coded caching system. The state-action value function is first approximated, based on which the policy is updated without needing ergodic search in the action space. Meanwhile, the parameters in the approximate expression are updated using the stochastic gradient descent (SGD) method.

The basic idea of function approximation here is to denote the state-action value function $Q(\mathbf{x},\mathbf{a})$ using a parameterized function approximator $\widehat{Q}(\mathbf{x},\mathbf{a}), \forall\mathbf{x},\mathbf{a}$, by taking into consideration the goal of the practical caching design. Since the state-action value function is a cumulation of the discounted rewards, inspired by (\ref{eqn_reward}), it can be approximated as
\setlength{\arraycolsep}{0.0em}
\begin{eqnarray}
\label{Q_approximate}
\widehat{Q}\left(\mathbf{x}(t),\mathbf{a}(t)\right)&&=\beta-\omega_{1}\sum_{i=1}^{C}\eta_{i}\theta_{i}(t)(1-da_{i}(t))u(1-da_{i}(t))\nonumber\\
&&-\omega_{2}\sum_{i=1}^{C}\xi_{i}(a_{i}(t)\!-\!a_{i}(t\!-\!1))u(a_{i}(t)-a_{i}(t\!-\!1)),\nonumber\\
&&
\end{eqnarray}
where $u(\cdot)$ is the unit step function that is equal to 1 if the value in the parentheses is equal or greater than 0 and 0 otherwise, $\beta$, $\eta_{i}$, and $\xi_{i}$ are unknown parameters, $\omega_{1}$ and $\omega_{2}$ represent weights of the corresponding parts, which are generally set as $\omega_{1}\gg\omega_{2}$ because the first part refers to the reduplicative requests of users during peak hours while the second part represents the non-repetitive requests of the $p$ SBSs during off-peak hours.

Instead of comparing the state-action values under all actions in each iteration in Algorithm 1, the caching action under the current state, $\mathbf{x}(t)$, can be obtained from the specific expression of the approximated value function and written as
\setlength{\arraycolsep}{0.0em}
\begin{eqnarray}
\label{action_chosen}
\mathbf{a}(t)&&=\arg\max\limits_{\mathbf{a}}\widehat{Q}\left(\mathbf{x}(t),\mathbf{a}\right)\nonumber\\
&&=\arg\max\limits_{a_{i},i\in\mathcal{C}}\biggl[\beta-\omega_{1}\sum_{i=1}^{C}\eta_{i}\theta_{i}(t)(1-da_{i})u(1-da_{i})\nonumber\\
&&\quad-\omega_{2}\sum_{i=1}^{C}\xi_{i}(a_{i}-a_{i}(t-1))u(a_{i}-a_{i}(t-1))\biggr].
\end{eqnarray}
Since $\omega_{1}$ is much larger than $\omega_{2}$, the alternative action can be obtained by omitting $\omega_{2}\sum_{i=1}^{C}\xi_{i}(a_{i}-a_{i}(t-1))u(a_{i}-a_{i}(t-1))$ as
\begin{equation}
\label{approxi_action_chosen}
\begin{aligned}
\hat{\mathbf{a}}(t)=\arg\max\limits_{a_{i},i\in\mathcal{C}}\biggl[\beta-\omega_{1}\sum_{i=1}^{C}\eta_{i}\theta_{i}(t)(1-da_{i})u(1-da_{i})\biggr].
\end{aligned}
\end{equation}
The procedure of solving (\ref{approxi_action_chosen}) is summarized as follows.

1) Determine the largest element of the action vector according to
\begin{equation}
\label{lmax_condition2}
\begin{aligned}
l_{\max}=\lceil\frac{L}{d}\rceil,
\end{aligned}
\end{equation}
where $l_{\max}\in\mathbb{N}^{+}$ denotes the numerator of the largest element.

2) Assume the number of contents corresponding to the caching fraction $\frac{i}{L}$ is $z_i$ for $i=1,2,\ldots,l_{\max}$ and $z_i$ is computed as
\begin{equation}
z_{i}=
\begin{cases}
\lfloor\frac{KL}{l_{\max}}\rfloor & i=l_{\max}\\
\lfloor\frac{KL-\sum_{j=1}^{l_{\max}-i}(i+j)z_{i+j}}{i}\rfloor & i=1,2,\ldots,l_{\max}-1.
\end{cases}
\end{equation}

3) Sort all the coefficients $\eta_{i}\theta_{i}(t)$, $i=1,2,\ldots,C$, as a vector in decending order and the $j$th element is $\eta_{h_{j}}\theta_{h_{j}}(t)$ corresponding to the $h_{j}$th content before sorting for $j=1,2,\ldots,C$. First, we roughly assign a value to each element of $\mathbf{\hat{a}}(t)$, which is given by
\begin{equation}
\hat{a}_{h_{j}}(t)=
\begin{cases}
\,\,\,\frac{l_{\max}}{L} & j=1,2,\ldots,z_{l_{\max}}\\
\frac{l_{\max}-1}{L} & j=z_{l_{\max}}\!\!+\!\!1,z_{l_{\max}}\!\!+\!\!2,\ldots,z_{l_{\max}}\!\!+\!\!z_{l_{\max}-1}\\
\quad\,\vdots\\
\,\,\,\,\,\frac{1}{L} & j=\!\!\!\sum\limits_{v=0}^{l_{\max}-2}\!\!z_{l_{\max}\!-\!v}\!\!+\!\!1,\ldots,\!\sum\limits_{v=0}^{l_{\max}-2}\!z_{l_{\max}\!-\!v}\!\!+\!\!z_{1}\\
\,\,\,\,\,\,0 & \textrm{others}.
\end{cases}
\end{equation}
Next, fine tune $\hat{a}_{h_{j}}(t)$ for $j=1,2,\ldots,z_{l_{\max}}$ with $1-d\cdot\frac{l_{\max}}{L}<0$. Starting with $j=z_{l_{\max}}$, repeat the following steps until $j=1$. For $j^{'}={z_{l_{\max}}+z_{l_{\max}-1}+1},\ldots,C$, find the minimum value of $j^{'}$, i.e. $\hat{j^{'}}$, satisfying
\setlength{\arraycolsep}{0.0em}
\begin{eqnarray}
\!\!\left(1-d\cdot\frac{l_{\max}-1}{L}\right)\cdot\eta_{h_{j}}\theta_{h_{j}}(t)<d\cdot\frac{1}{L}\cdot\eta_{h_{\hat{j^{'}}}}\theta_{h_{\hat{j^{'}}}}(t).
\end{eqnarray}
Then adjust $\hat{a}_{h_{j}}(t)$ by subtracting $\frac{1}{L}$ and $\hat{a}_{h_{\hat{j^{'}}}}(t)$ by adding $\frac{1}{L}$.

The objective is to make the parameterized state-action value function $\widehat{Q}\left(\mathbf{x},\mathbf{a}\right)$ and the real $Q\left(\mathbf{x},\mathbf{a}\right)$ as close as possible. The loss function is defined as
\begin{equation}
\begin{aligned}
\label{Q_loss}
Loss(\beta,\boldsymbol{\eta},\boldsymbol{\xi})=\mathbb{E}\left[(Q(\mathbf{x},\mathbf{a})-\widehat{Q}(\mathbf{x},\mathbf{a}))^{2}\right],
\end{aligned}
\end{equation}
where $\boldsymbol{\eta}\!=\![\eta_{1},\eta_{2},\ldots,\eta_{C}]$ and $\boldsymbol{\xi}\!=\![\xi_{1},\xi_{2},\ldots,\xi_{C}]$. $Q(\mathbf{x},\mathbf{a})$ is unknown and can be replaced by the currently estimated value function, $R_{\mathbf{x}\rightarrow\mathbf{x}^{'}}^{\mathbf{a}}\!+\!\gamma\widehat{Q}\left(\mathbf{x}^{'},\mathbf{\tilde{a}}\right)$, resorting to the temporal difference learning. The estimated caching action under state $\mathbf{x}^{'}$, which is denoted by $\mathbf{\tilde{a}}\!=\!\arg\max\limits_{\mathbf{a}^{'}}\widehat{Q}\left(\mathbf{x}^{'},\mathbf{a}^{'}\right)$, can also be obtained following the same procedure as the above. Based on each sample $\left(\mathbf{x}(t),\mathbf{a}(t)\right)$, the parameters, $\{\beta,\boldsymbol{\eta},\boldsymbol{\xi}\}$, are updated resorting to the SGD method to minimize the loss function as
\begin{subequations}
\label{para_update}
\begin{align}
\beta^{\textrm{cu}}&=\beta^{\textrm{pr}}-\delta\frac{\partial Loss}{\partial\beta}\nonumber\\
&=\beta^{\textrm{pr}}+2\delta\left(Q(\mathbf{x}(t),\mathbf{a}(t))-\widehat{Q}\left(\mathbf{x}(t),\mathbf{a}(t)\right)\right),\\
\eta_{i}^{\textrm{cu}}&=\eta_{i}^{\textrm{pr}}-\delta\frac{\partial Loss}{\partial\eta_{i}}\nonumber\\
&=\eta_{i}^{\textrm{pr}}-2\delta\omega_{1}\theta_{i}(t)\left(Q(\mathbf{x}(t),\mathbf{a}(t))-\widehat{Q}\left(\mathbf{x}(t),\mathbf{a}(t)\right)\right)\nonumber\\
&\quad\times(1-da_{i}(t))u\left(1-da_{i}(t)\right),\\
\xi_{i}^{\textrm{cu}}&=\xi_{i}^{\textrm{pr}}-\delta\frac{\partial Loss}{\partial\xi_{i}}\nonumber\\
&=\xi_{i}^{\textrm{pr}}-2\delta\omega_{2}\left(Q(\mathbf{x}(t),\mathbf{a}(t))-\widehat{Q}\left(\mathbf{x}(t),\mathbf{a}(t)\right)\right)\nonumber\\
&\quad\times(a_{i}(t)-a_{i}(t-1))u\left(a_{i}(t)-a_{i}(t-1)\right),
\end{align}
\end{subequations}
where $\beta^{\textrm{cu}}$, $\eta_{i}^{\textrm{cu}}$, and $\xi_{i}^{\textrm{cu}}$ denote the parameters in the current time slot, $\beta^{\textrm{pr}}$, $\eta_{i}^{\textrm{pr}}$, and $\xi_{i}^{\textrm{pr}}$ represent the parameters in the previous time slot, and $\delta$ is the step size.

Based on the procedure to find the caching action and the method to update the parameters described above, the value function approximation based cooperative coded caching strategy is summarized in Algorithm 2 in detail.

\begin{algorithm}
\label{alg:valuefunction}
\caption{Value Function Approximation based Cooperative Coded Caching Strategy}
{\bf Input:}
environment simulator, the structure of the value function approximator $\hat{Q}(\mathbf{x},\mathbf{a})$\\
{\bf Output:}
the value function approximator $\hat{Q}(\mathbf{x},\mathbf{a})$\\
{\bf Procedure:}
\begin{algorithmic}[1]
\STATE Set the starting time slot $t=t_{0}$, the starting state $\mathbf{x}(t)=\mathbf{x}_{0}$, the discount factor $\gamma$, the step size $\delta$, the exploration probability $\epsilon$, the initial value of parameters: $\beta^{\textrm{pr}}=0$, $\boldsymbol{\eta}^{\textrm{pr}}=\boldsymbol{0}$, $\boldsymbol{\xi}^{\textrm{pr}}=\boldsymbol{0}$
\LOOP
\STATE (\emph{Selection of Cooperative Caching Action}): Choose a caching action based on $\varepsilon$-greedy approach during off-peak hours

\textrm{w.p.}\,$1-\varepsilon$

solve $\mathbf{a}(t)=\arg\max\limits_{\mathbf{a}}\hat{Q}\left(\mathbf{x}(t),\mathbf{a}\right)$ based on (\ref{approxi_action_chosen}) and the procedure 1), 2), 3)

\textrm{w.p.}\,$\varepsilon$

choose an action randomly that meets the constraints $\sum_{i=1}^{C}a_{i}(t)=K$ and $l_{\max}=\lceil\frac{L}{d}\rceil$
\STATE (\emph{MDS Coding}): MBS codes the contents with $a_{c}(t)\neq0$ and $a_{c}(t)\geq a_{c}(t-1)$ based on MDS coding during off-peak hours
\STATE (\emph{Coded Packets Delivery}): MBS sends $p$ different packets, each containing $(a_{c}(t)-a_{c}(t-1))B$ bits of the $c$th coded content, to $p$ SBSs correspondingly for caching during off-peak hours
\STATE (\emph{Cooperative Transmission}): User $g\in\mathcal{G}_{s}$ selects $d$ SBSs to form $\mathcal{S}_{g}^{d}$ and requests contents during peak hours of time slot $t+1$
\STATE (\emph{New State Observation}): MBS observes $\mathbf{N}(t+1)$, computes $\boldsymbol{\theta}(t+1)$ according to (\ref{eqn_theta}) during off-peak hours, and sets the new state $\mathbf{x}(t+1)=\left[\boldsymbol{\theta}(t+1),\mathbf{a}(t)\right]$
\STATE (\emph{Reward Feedback}): Environment feeds back a reward computed as (\ref{eqn_reward})
\STATE (\emph{Estimation of Cooperative Caching Action}): Estimate a caching action under the state $\mathbf{x}(t+1)$ by solving $\mathbf{\tilde{a}}(t+1)=\arg\max\limits_{\mathbf{a}^{'}}\hat{Q}\left(\mathbf{x}(t+1),\mathbf{a}^{'}\right)$ based on (\ref{approxi_action_chosen}) and the procedure 1), 2), 3)
\STATE (\emph{Parameters Updating}): Update the parameters $\{\beta,\boldsymbol{\eta},\boldsymbol{\xi}\}$ based on (\ref{para_update})
\STATE $t=t+1$
\ENDLOOP
\RETURN The value function approximator $\hat{Q}(\mathbf{x},\mathbf{a})$
\end{algorithmic}
\end{algorithm}

\section{Simulation Results}

In this section, simulation results are presented to verify the proposed RL based cooperative coded caching strategies for small-scale and large-scale systems. Unless stated otherwise, the Zipf-like distribution is used to generate the user requests, based on which the content popularity profiles can be computed according to (\ref{eqn_theta}). The following four baselines are used for performance comparison.

\begin{itemize}
  \item \textbf{Baseline 1} (Value Iteration based Optimal Caching): The state value function, $V_{\gamma}^{\pi}(\mathbf{x})$, denotes the expected cumulative discounted rewards following the policy $\pi$ under the current state $\mathbf{x}$. The state value function of the optimal policy is expressed as $V_{\gamma}^{\pi^{*}}(\mathbf{x})=\max\limits_{\mathbf{a}\in\mathcal{A}}Q_{\gamma}^{\pi^{*}}(\mathbf{x},\mathbf{a})$. Under the assumption that the agent knows the transition tuple described in III-A, the state value function for each $\mathbf{x}$ is updated according to $V(\mathbf{x})=\max\limits_{\mathbf{a}\in\mathcal{A}}\sum_{\mathbf{x}^{'}\in\mathcal{X}}P_{\mathbf{x}\rightarrow\mathbf{x}^{'}}^{\mathbf{a}}\\
      \times\left(R_{\mathbf{x}\rightarrow\mathbf{x}^{'}}^{\mathbf{a}}+\gamma V(\mathbf{x}^{'})\right)$ until converging to $V_{\gamma}^{\pi^{*}}(\mathbf{x})$. Then the optimal policy is obtained as $\pi^{*}(\mathbf{x})=\arg\max\limits_{\mathbf{a}\in\mathcal{A}}Q_{\gamma}^{\pi^{*}}(\mathbf{x},\mathbf{a})$.
  \item \textbf{Baseline 2} (Most Popular based Cooperative Caching (MPCC)): According to $\boldsymbol{\theta}(t)$ at time slot $t$, the agent selects the most popular contents to cache incorporating the MDS coding while the interactions with the environment are not considered, i.e., without RL. It can be seen how this baseline scheme caches contents based on $\boldsymbol{\theta}(t)$ from Fig.~\ref{popularity_action}.
  \item \textbf{Baseline 3} (RL based Non-cooperative Caching (RL-NC)): The agent determines the caching strategy based on RL while the cooperation among SBSs is not considered, which corresponds to a special case when $d=1$ in our proposed algorithms. In this case, full contents are cached at the SBSs.
  \item \textbf{Baseline 4} (RL based Uncoded Cooperative Caching (RL-UCC)): The agent determines the caching strategy based on RL with SBS cooperation while the fragments stored at the SBSs are uncoded. Specifically, each SBS randomly stores the fragments with the corresponding size according to the action made by the agent.
\end{itemize}

By comparing the proposed algorithms with Baselines 2, 3, and 4, respectively, the effects of RL, the combination of MDS coding and SBS cooperation, and MDS coding employed in the proposed algorithms will be clear.

To measure and compare various schemes, define the direct SBS-serving ratio as
\begin{equation}
\begin{aligned}
\label{Ratio}
\rho=\frac{N_{\textrm{SBS}}}{N_{\textrm{total}}},
\end{aligned}
\end{equation}
where $N_{\textrm{SBS}}$ is the traffic load served directly by the SBSs and $N_{\textrm{total}}$ represents the total traffic load.\footnote{In fact, $N_{\textrm{SBS}}$ is also the reward in our formulated RL problem and $N_{\textrm{total}}$ is also the total number of content requests.}

\vspace{-0.2cm}
\subsection{Performance in Small-Scale System}

In the small-scale system, two content popularity profiles, $\boldsymbol{\mathbf{\theta}}_{1}$ and $\boldsymbol{\mathbf{\theta}}_{2}$, are considered and computed according to (\ref{eqn_theta}), where the user requests are generated based on the Zipf-like distribution \cite{S. Gitzenis} with $\alpha_1$ and $\alpha_2$ denoting the corresponding skewness. Using the case with the starting state $[\boldsymbol{\mathbf{\theta}}_{1},\mathbf{a}_{i}]$ and the selected action $\mathbf{a}_{j}$ as an example, the transition probabilities are chosen randomly from $[0,1]$ and subject to the following constraints
\begin{subequations}
\begin{align}
&P_{\left[\boldsymbol{\theta}_{1},\mathbf{a}_{i}\right]\rightarrow\left[\boldsymbol{\theta}_{1},\mathbf{a}_{k}\right]}^{\mathbf{a}_{j}}+P_{\left[\boldsymbol{\theta}_{1},\mathbf{a}_{i}\right]\rightarrow\left[\boldsymbol{\theta}_{2},\mathbf{a}_{k}\right]}^{\mathbf{a}_{j}}=1\;\;\;k=j,\\
&P_{\left[\boldsymbol{\theta}_{1},\mathbf{a}_{i}\right]\rightarrow[\boldsymbol{\theta}_{1},\mathbf{a}_{k}]}^{\mathbf{a}_{j}}=0\;\;\;k\neq j,\\
&P_{\left[\boldsymbol{\theta}_{1},\mathbf{a}_{i}\right]\rightarrow[\boldsymbol{\theta}_{2},\mathbf{a}_{k}]}^{\mathbf{a}_{j}}=0\;\;\;k\neq j.
\end{align}
\end{subequations}

Major simulation parameters in the small-scale system are listed in Table \ref{Sim_para_small}.

\begin{table}
\small
  \centering
  \caption{Simulation Parameters in Small-Scale System}\label{Sim_para_small}
  \label{tab:parameters}
  \begin{tabular}{c|c}
  \hline
  Simulation Parameter & Setting Value\\
  \hline
  $p$ & 20\\
  \hline
  $C$ & 10\\
  \hline
  $K$ & [1,2,3,4]\\
  \hline
  $d$ & 2\\
  \hline
  $L$ & 3\\
  \hline
  $\gamma$ & 0.9\\
  \hline
  $\lambda$ & 0.6\\
  \hline
  $\alpha_{1}$ & 1.36\\
  \hline
  $\alpha_{2}$ & 2.3\\
  \hline
  $\epsilon$ for exploration stage & 0.1\\
  \hline
  $\epsilon$ for exploitation stage & 0\\
  \hline
  \end{tabular}
\end{table}

\begin{figure}[!t]
\centering
\includegraphics[trim=15 15 0 10, width=3.4in]{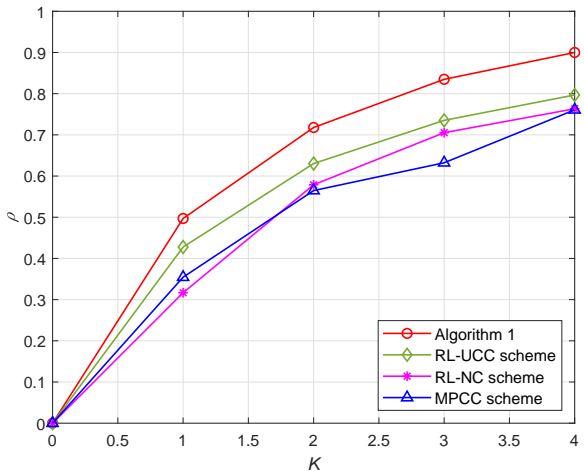}
\caption{The direct SBS-serving ratio for Algorithm 1, the MPCC scheme, the RL-NC scheme, and the RL-UCC scheme in the small-scale system.}\label{cachesize_small}
\end{figure}

To compare caching efficiency, Fig.~\ref{cachesize_small} plots the direct SBS-serving ratio, $\rho$, against growing cache size, $K$, for Algorithm 1, the MPCC scheme, the RL-NC scheme, and the RL-UCC scheme in the small-scale system. From the figure, the performance improves with cache size for all caching schemes and the proposed Algorithm 1 performs better than the other three baselines. By comparing Algorithm 1 and MPCC, we can see the significant advantage of using RL to make caching decision. Algorithm 1 also outperforms RL-NC resorting to the combination of MDS coding and SBS cooperation. The RL-UCC scheme achieves better performance than MPCC and RL-NC but still has a significant gap to Algorithm 1 since the MDS coding based caching considered in Algorithm 1 can utilize the limited cache storage more efficiently than the uncoded random caching used in RL-UCC. In more details, each SBS caches the individual coded packet irrelevant to each other in the MDS coding based caching, which guarantees the successful content recovery so long as the user collects a certain number of packets with the total size no less than the size of the original content from the SBSs. In the uncoded random caching, the SBSs cache the uncoded fragments of each content randomly and thus the user may receive repetitive fragments from the SBSs, in which case the performance of RL-UCC is degraded. For example, Algorithm 1 achieves performance gains of about $0.14$, $0.14$, and $0.1$ compared to MPCC, RL-NC, and RL-UCC, respectively, when the cache size $K=4$. We can also find that the effect of RL is larger than the MDS coding in the small-scale system by comparing MPCC and RL-UCC. To converge to the corresponding $\rho$, Algorithm 1, RL-UCC, and RL-NC consume $375.8$ seconds, $372.3$ seconds, and $80.5$ seconds, respectively, when $K=1$. The corresponding time increases to $1034.6$ seconds, $1300.9$ seconds, and $110.2$ seconds, respectively, when $K=2$. So the consumed time for convergence is dependent on the size of the action space. In addition, SBS cooperation will enlarge the action space, which is the reason that Algorithm 1 and RL-UCC use more time than RL-NC.

\begin{figure}[!t]
\centering
\includegraphics[trim=15 15 0 10, width=3.4in]{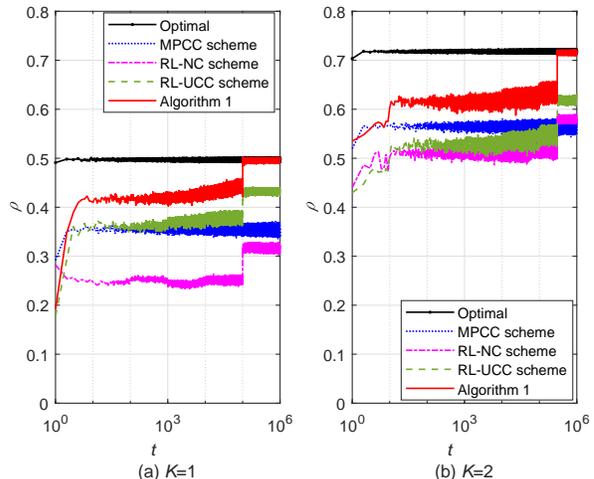}
\caption{The convergence performance of Algorithm 1 and baseline schemes with $K=1$ and $2$.}\label{convergence_Qlearning}
\end{figure}

To show the convergence behaviors of the proposed Algorithm 1 and the baseline schemes, Fig.~\ref{convergence_Qlearning} plots the direct SBS-serving ratio, $\rho$, with increasing time slot, $t$, with the cache size $K=1$ and $2$. From Fig.~\ref{convergence_Qlearning}(a), the performance of MPCC stabilizes at about $0.35$ fast. In contrast, Algorithm 1 achieves the increasing performance with $t$ in the exploration stage with $t<10^5$ and finally converges to the optimal value, $\rho=0.5$, by setting $\epsilon=0$ in the exploitation stage starting from $t=10^5$, demonstrating the effectiveness of caching policy learnt through interactions with the environment.\footnote{It is noted that the switch point, $t=10^5$, from the exploration stage to the exploitation stage is determined according to the simulation trials. Setting $\epsilon$ to $0$ untimely will lead to insufficient exploration in action space and consequently degenerate the converged performance.} The performance of RL-UCC has the similar trend but suffers a degradation due to using the less efficient uncoded random caching. More performance loss is incurred for RL-NC since both SBS cooperation and MDS coding are not utilized. The action space enlarges with $K=2$, based on which we allocate more time slots for the exploration stage and set $\epsilon=0$ at $t=3\times10^{5}$ in Fig.~\ref{convergence_Qlearning}(b). Compared to Fig.~\ref{convergence_Qlearning}(a), Fig.~\ref{convergence_Qlearning}(b) shows the similar convergence behavior and increased converged performance due to the larger cache size for each scheme.

\begin{figure}[!t]
\centering
\includegraphics[trim=15 15 0 10, width=3.4in]{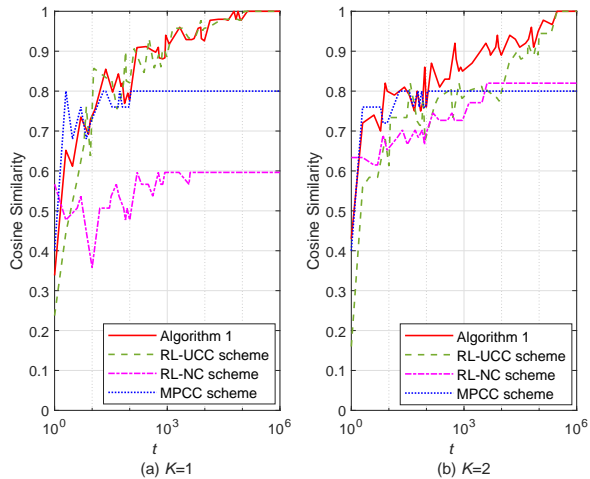}
\caption{Cosine similarity between the optimal action vector and those of Algorithm 1, the MPCC scheme, the RL-NC scheme, and the RL-UCC scheme with $K=1$ and $2$.}\label{Similarity}
\end{figure}

To provide an intuitive explanation about the performance gain achieved by Algorithm 1, we plot the curve of cosine similarity with $K=1$ and $2$ in Fig.~\ref{Similarity}. The cosine similarity is defined as $\cos(\phi_{\mathbf{a}_{\textrm{opt}},\Delta})=\frac{\langle{\mathbf{a}}_{\textrm{opt}}^{T},\:\Delta^{T}\rangle}{\|\mathbf{a}_{\textrm{opt}}\|\cdot\|\Delta\|}$ with $\langle\cdot,\cdot\rangle$, $\|\cdot\|$, and $(\cdot)^{T}$ representing the inner product, Euclidean norm, and transpose, where $\mathbf{a}_{\textrm{opt}}$, $\Delta$, and $\phi_{\mathbf{a}_{\textrm{opt}},\Delta}$ denote the optimal action vector, the corresponding action vector of Algorithm 1, MPCC, RL-NC, and RL-UCC, and the angle between $\mathbf{a}_{\textrm{opt}}$ and $\Delta$, respectively. From Fig.~\ref{Similarity}(a), the similarity curves of Algorithm 1 and the other three baseline schemes are on the rise as $t$ increases until reaching their respective convergence values. The curves of MPCC and RL-NC stabilize at about $0.8$ and $0.6$, respectively, while the curves of Algorithm 1 and RL-UCC finally converge to $1$. That is, the superiority of Algorithm 1 to MPCC and RL-NC mainly owes to the effect of RL and the effect of the combination of SBS cooperation and MDS coding, respectively. Both Algorithm 1 and RL-UCC are able to learn the optimal action, which demonstrates that Algorithm 1 outperforms RL-UCC resorting to the advantage of MDS coding based caching over the uncoded random caching. For Fig.~\ref{Similarity}(b) with $K=2$, Algorithm 1 and RL-UCC can still find the optimal action while the similarity performance of RL-NC and MPCC finally converges to $0.82$ and $0.8$, respectively. Through Fig.~\ref{Similarity}, the effects of RL, SBS cooperation, and MDS coding are revealed clearly.

\begin{figure}
\centering
\subfigure[Popularity profiles of $3$ successive time slots.]{
\centering
\includegraphics[trim=0 55 0 10, height=1.9in,width=3.1in]{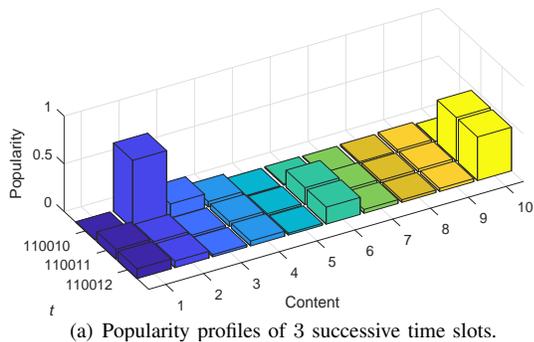}
}
\vspace{-0.3cm}

\subfigure[Actions of different schemes at the end of the first time slot.]{
\centering
\includegraphics[trim=0 15 0 3, height=2.2in,width=3.3in]{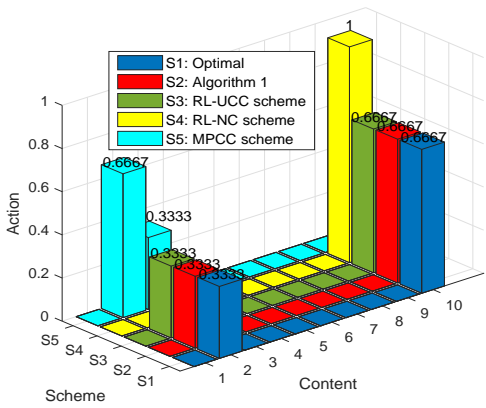}
}
\caption{A toy example: Popularity profiles and corresponding actions with $K=1$.}\label{popularity_action}
\end{figure}

To get more insights, we use a toy example to show how they take action to match the content popularity in Fig.~\ref{popularity_action}. Specifically, Fig.~\ref{popularity_action}(a) plots the popularity profiles of three successive time slots and Fig.~\ref{popularity_action}(b) plots the action of each scheme at the end of the first time slot $110010$. From the figure, the MPCC scheme simply selects the action based on the popularity profile of time slot $110010$, which obviously mismatches the upcoming popularity profile. RL-NC, RL-UCC, and Algorithm 1 all select action based on the long-term performance. However, RL-NC only caches full content and inevitably causes performance loss. In contrast, RL-UCC and Algorithm 1 smartly allocate the action to the most popular contents in all popularity profiles, which is the optimal action. This again reveals that the uncoded random caching used by RL-UCC wastes the limited cache storage and thus performs worse than Algorithm 1 even though they have learned very similar caching policy.

\vspace{-0.2cm}
\subsection{Performance in Large-Scale System}

In the large-scale system, ten content popularity profiles are considered and computed according to (\ref{eqn_theta}), where the user requests are generated based on Zipf-like distribution \cite{S. Gitzenis} with $\alpha_i=1.2+0.2(i-1)$, for $i=1,\ldots,10$, denoting the skewness of the $i$th user request distribution. And the transition probabilities are generated randomly from $[0,1]$ under the constraints $\sum_{j=1}^{10}P_{\left[\boldsymbol{\theta}_{i},\mathbf{a}\right]\rightarrow\left[\boldsymbol{\theta}_{j},\mathbf{a}^{'}\right]}^{\mathbf{a}^{'}}=1$, for $ i=1,\ldots,10$, where $\mathbf{a}^{'}$ is the chosen action under the state $\left[\boldsymbol{\theta}_{i},\mathbf{a}\right]$.
Major simulation parameters in the large-scale system are listed in Table \ref{Sim_para_large}.

\begin{table}
\small
  \centering
  \caption{Simulation Parameters in Large-Scale System}\label{Sim_para_large}
  \label{tab:parameters}
  \begin{tabular}{c|c}
  \hline
  Simulation Parameter & Setting Value\\
  \hline
  $p$ & 50\\
  \hline
  $C$ & 100\\
  \hline
  $K$ & [5,10,15,20]\\
  \hline
  $d$ & 3\\
  \hline
  $L$ & 6\\
  \hline
  $\gamma$ & 0.9\\
  \hline
  \makecell{$\delta$ for Algorithm 2\\ and RL-UCC scheme} & \makecell{[0.01,0.01,0.1,0.5] \\
  corresponds to $K$}\\
  \hline
  $\delta$ for RL-NC scheme & \makecell{[0.01,0.01,0.8,0.8] \\
  corresponds to $K$}\\
  \hline
  $\epsilon$ for exploration stage & 0.2\\
  \hline
  $\epsilon$ for exploitation stage & 0\\
  \hline

  \end{tabular}
\end{table}

\begin{figure}[!t]
\centering
\includegraphics[trim=15 10 0 15, width=3.4in]{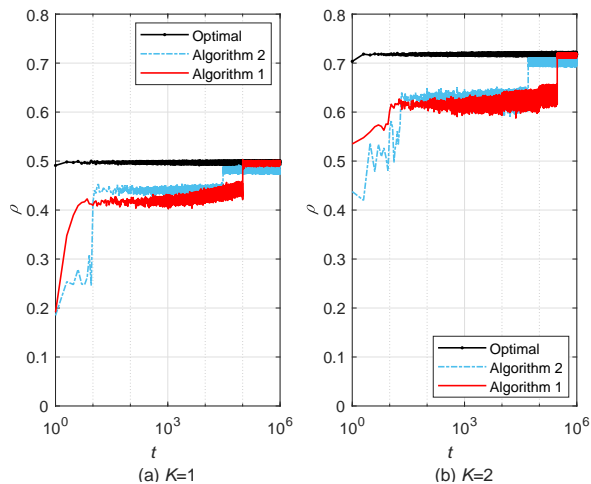}
\caption{The convergence performance of Algorithm 2 with $K=1$ and $2$ in the small-scale system.}\label{convergence_VFA}
\end{figure}

Fig.~\ref{convergence_VFA} shows the convergence performance of Algorithm 2, where the optimal scheme and Algorithm 1 are also plotted to demonstrate the near-optimality and fast convergence of Algorithm 2. All the schemes in this figure still use the parameters listed in Table~\ref{Sim_para_small} since the optimal scheme and Algorithm 1 can be obtained only in the small-scale system due to high computational complexity. Although Algorithm 2 is designed for the large-scale system, it is also feasible in the small-scale system. In Fig.~\ref{convergence_VFA}(a) and (b), Algorithm 2 employs value function approximation to avoid the infinite execution for each state-action pair and the onerous ergodic search over the action space. As a result, Algorithm 2 converges faster than Algorithm 1 at the cost of only converging to the near-optimal point. Nevertheless, Algorithm 2 can be flexibly applied to the large-scale system, in which case Algorithm 1 is infeasible. From the figure, the exploitation-only performance of Algorithm 1 is better than that of Algorithm 2 during the exploitation stage ($\epsilon=0$), which is because it finds the optimal policy through exhaustive search while Algorithm 2 only converges to a sub-optimal solution with the approximated value function. However, during the exploration stage ($\epsilon>0$), the exploration in the action space brings some uncertainties to the exploration-exploitation performance. As a result, Algorithm 1 may achieve the performance similar to Algorithm 2 or outperform Algorithm 2 at the switch point.

\begin{figure}[!t]
\centering
\includegraphics[trim=15 20 0 25, width=3.4in]{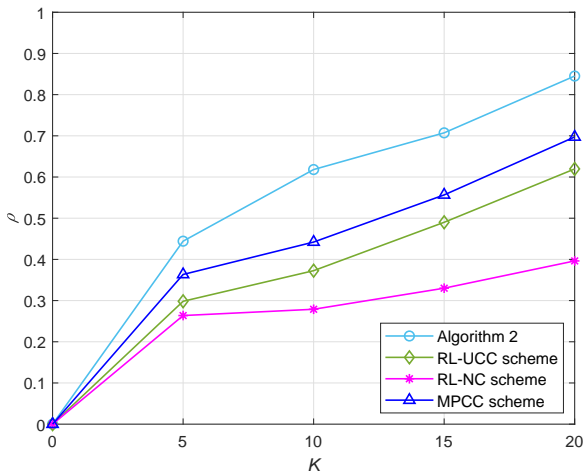}
\caption{The direct SBS-serving ratio for Algorithm 2, the MPCC scheme, the RL-NC scheme, and the RL-UCC scheme in the large-scale system.}\label{large_scale_cachesize}
\end{figure}

Fig.~\ref{large_scale_cachesize} plots the direct SBS-serving ratio, $\rho$, versus the cache size, $K$, for Algorithm 2, the MPCC scheme, the RL-NC scheme, and the RL-UCC scheme in the large-scale system, where Algorithm 2 outperforms the other three baselines more significantly than in the small-scale system. When the cache size $K=15$, Algorithm 2 achieves the direct SBS-serving ratio $\rho=0.71$ while the corresponding performance of MPCC, RL-NC, and RL-UCC are $0.55$, $0.33$, and $0.49$ respectively, which means that Algorithm 2 can achieve the considerable performance even with limited cache size. The effectiveness of Algorithm 2 remedies the infeasibility of Algorithm 1 in the large-scale system and makes our proposed RL based cooperative coded caching scheme applicable to various practical systems. In addition, the superiority of MPCC over RL-UCC demonstrates that the uncoded random caching causes more significant performance loss and the MDS coding based caching becomes more important in the large-scale system.

\begin{figure}[!t]
\centering
\includegraphics[trim=15 20 0 20, width=3.4in]{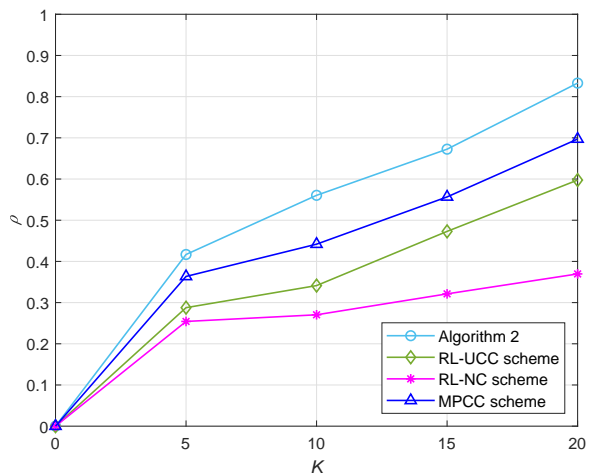}
\caption{The direct SBS-serving ratio for Algorithm 2, the MPCC scheme, the RL-NC scheme, and the RL-UCC scheme in the large-scale system under SNM based user requests.}\label{large_scale_locality}
\end{figure}

To verify the robustness of Algorithm 2, we introduce the temporal locality to the user requests based on the shot noise model (SNM) according to \cite{S. Traverso}, where a small amount of contents are popular temporarily. Fig.~\ref{large_scale_locality} plots the direct SBS-serving ratio for Algorithm 2 and the three baseline schemes in the large-scale system under the SNM based user requests. Similar to Fig.~\ref{large_scale_cachesize}, Algorithm 2 still achieves remarkable advantage compared to the baseline schemes. The SNM based user requests with temporal locality only cause slight performance loss for Algorithm 2.

\begin{figure}[!t]
\centering
\includegraphics[trim=15 20 0 25, width=3.4in]{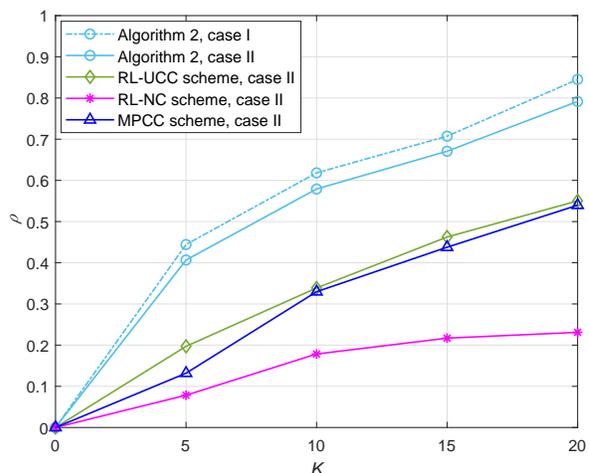}
\caption{The robustness of Algorithms 2 in the large-scale system.}\label{large_scale_robustness}
\end{figure}

To further demonstrate the robustness of Algorithm 2 in the time-varying environment, Fig.~\ref{large_scale_robustness} plots the performance of Algorithm 2 and the baseline schemes in two cases, where case I uses the environment set in the first paragraph of Section V.B while case II uses another set of content popularity candidates and transition probabilities. From Fig.~\ref{large_scale_robustness}, the environment changing only leads to a minor performance loss for Algorithm 2. When facing case II with the changed environment, Algorithm 2 achieves the more significant advantage over other baseline schemes compared to Fig.~\ref{large_scale_cachesize}.

\subsection{Complexity Comparison}

\begin{table}[!t]
\centering
\caption{Complexity Comparison}
\label{table_complexity}
\begin{tabular}
{c|c|c}
\hline
 Scheme & \makecell{Small-scale system} & \makecell{Large-scale system} \\
\hline
 MPCC & \makecell{$\mathcal{O}(C^2$\\$+CB_{c}^{\textrm{MDS}}\log B_{c}^{\textrm{MDS}})$} & \makecell{$\mathcal{O}(C^2$\\$+CB_{c}^{\textrm{MDS}}\log B_{c}^{\textrm{MDS}})$} \\
\hline
 RL-NC & $\mathcal{O}(T_{\textrm{RN}}|\mathcal{A}_{\textrm{RN}}|)$ & $\mathcal{O}(T_{\textrm{RN}}C)$ \\
\hline
 RL-UCC & $\mathcal{O}(T_{\textrm{RU}}|\mathcal{A}_{\textrm{RU}}|)$ & $\mathcal{O}(T_{\textrm{RU}}C)$ \\
\hline
 Algorithm 1 & \makecell{$\mathcal{O}(T_{1}(|\mathcal{A}_{\textrm{1}}|$\\$+CB_{c}^{\textrm{MDS}}\log B_{c}^{\textrm{MDS}}))$} & - \\
\hline
 Algorithm 2 & $\mathcal{O}(T_{2}CB_{c}^{\textrm{MDS}}\log B_{c}^{\textrm{MDS}})$ & $\mathcal{O}(T_{2}CB_{c}^{\textrm{MDS}}\log B_{c}^{\textrm{MDS}})$ \\
\hline
\end{tabular}
\end{table}

Table~\ref{table_complexity} lists the computational complexity of MPCC, RL-NC, RL-UCC, Algorithm 1, and Algorithm 2 when achieving the converged performance in small- and large-scale systems. $T_{\textrm{RN}}$, $T_{\textrm{RU}}$, $T_{1}$, and $\mathcal{A}_{\textrm{RN}}$, $\mathcal{A}_{\textrm{RU}}$, $\mathcal{A}_{1}$ denote the total steps for convergence and action space of RL-NC, RL-UCC, and Algorithm 1, respectively. $T_{2}$ denotes the total steps for convergence of Algorithm 2. MPCC makes decision simply based on the current popularity profile without needing iteration and thus its complexity does not include the number of steps for convergence. RL-NC, RL-UCC, and Algorithm 1 in the small-scale system apply Q-learning with Q-table updating and the complexity, dependent on the size of the action space, will be prohibitively huge in the large-scale system. It can be seen that Algorithm 2 significantly reduces the complexity compared to Algorithm 1 and causes the complexity of only $\mathcal{O}(C)$ for parameter updating in each step. RL-NC and RL-UCC also use value function approximation similar to Algorithm 2 to reduce the complexity in the large-scale system. By using the comparable or moderately increased complexity, Algorithms 1 and 2 achieve the better performance than other baseline schemes.

For Algorithm 2, $T_{2}$ increases with the size of action space that depends on the number of contents, $C$, the action discretization level, $L$, and the cache size, $K$, while $B_{c}^{\textrm{MDS}}$ increases with the number of SBSs, $p$, according to (\ref{eqn_BMDS}). The complexity reveals the key system parameters impacting the capability of Algorithm 2. Although the complexity of Algorithm 2 increases with $p$, $C$, and $L$, it is still feasible so long as these parameters are not too large. When they grow to very large values, we can omit the contents that are barely requested, which account for the majority of all contents \cite{S. Traverso}. We can also reduce $L$ appropriately. With these solutions, Algorithm 2 is able to work in the very large-scale system at the cost of the acceptable performance loss.

\subsection{Impact of Action Discretization}

\begin{figure}[!t]
\centering
\includegraphics[trim=15 10 0 20, width=3.4in]{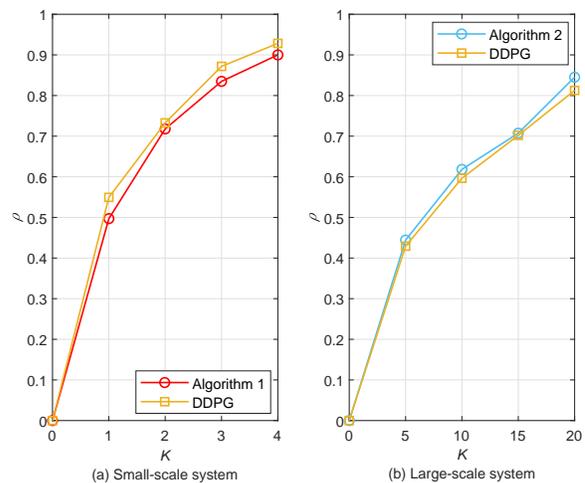}
\caption{Performance comparison between the proposed algorithms and DDPG.}\label{Algorithm12_DDPG}
\end{figure}

To reveal the impact of action discretization used in the proposed algorithms, we compare the deep deterministic policy gradient (DDPG) algorithm \cite{Y. Wei} that is able to output the continuous action with Algorithm 1 in the small-scale system and with Algorithm 2 in the large-scale system, respectively, in Fig.~\ref{Algorithm12_DDPG}. From Fig.~\ref{Algorithm12_DDPG}(a), Algorithm 1 achieves very similar performance to DDPG, indicating that the action discretization only causes a very limited loss.\footnote{Algorithm 1 is the optimal solution for the discrete action case and thus has a tiny performance gap to DDPG.} To achieve their respective performance at $K=1$, Algorithm 1 consumes only $375.8$ seconds while DDPG needs $1296.6$ seconds. DDPG needs much more time to find the appropriate continuous action, which however only brings a marginal performance improvement. For the large-scale case in Fig.~\ref{Algorithm12_DDPG}(b), Algorithm 2 slightly outperforms DDPG even if it discretizes the action space. This reveals that finding the appropriate action from a continuous space becomes more difficult for DDPG since it has the complicated network structures of actor and critic and needs to update much more parameters in each iteration. On the other hand, Algorithm 2 needs $565.3$ seconds to achieve its performance at $K=5$ while the time consumed by DDPG is $3411.9$ seconds. Algorithm 2 benefits from the insightful approximation for the state-action value function and the efficient action selection in each iteration. Therefore, action discretization is preferable in our considered caching design problem since it hardly impacts the performance but reduces the computational complexity significantly.

\section{Conclusion}

In this paper, we have studied the RL based cooperative coded caching strategy design for SBSs in UDNs. The practical time-variant popularity profile with the unknown evolution way is taken into consideration to formulate the MDS coding based cooperative caching design in the perspective of RL. The optimal solution is first developed by embedding the complicated cooperative MDS coding into Q-learning. For the large-scale system, we further propose an insightful approximation for the state-action value function heuristically along with an efficient action-selection approach, which is proved near-optimal but with significantly reduced complexity. The proposed RL based cooperative coded caching strategy has significant performance superiority and is applicable to practical systems with different dimensionalities and environments.

In the future, we can further consider that the SBSs use their respective caching policies in the multi-agent RL manner. For multi-agent RL, the agents impact each other in decision making and thus independent Q-learning based on own action and observation will lead to the nonstationary environment for each agent \cite{L. Liang_b}, which needs to be addressed in the distributed framework for caching policy design.



\ifCLASSOPTIONcaptionsoff
  \newpage
\fi

\end{document}